\documentclass[reprint, amsmath, amssymb, aps, prl, superscriptaddress]{revtex4-1}

\usepackage{graphicx}
\usepackage{dcolumn}
\usepackage{amsmath}
\usepackage{amsfonts}
\usepackage{epsfig}
\usepackage{bm}
\usepackage{color}
\usepackage{mathtools}

\begin{document}


\title{Biexciton Condensation in Electron-Hole-Doped Hubbard Bilayers -- \\ A Sign-Problem-Free Quantum Monte Carlo Study}

\author{Xu-Xin Huang}
\affiliation{Department of Applied Physics, Stanford University, Stanford, CA 94305, USA}
\affiliation{Stanford Institute for Materials and Energy Sciences, SLAC National Accelerator Laboratory and Stanford University, 2575 Sand Hill Road, Menlo Park, CA 94025, USA}
\author{Martin Claassen}
\affiliation{Center for Computational Quantum Physics, Flatiron Institute, Simons Foundation, 162 5th Ave., New York, NY 10010, USA}
\author{Edwin W. Huang}
\affiliation{Stanford Institute for Materials and Energy Sciences, SLAC National Accelerator Laboratory and Stanford University, 2575 Sand Hill Road, Menlo Park, CA 94025, USA}
\affiliation{Department of Physics, Stanford University, Stanford, CA 94305, USA}
\author{Brian Moritz}
\affiliation{Stanford Institute for Materials and Energy Sciences, SLAC National Accelerator Laboratory and Stanford University, 2575 Sand Hill Road, Menlo Park, CA 94025, USA}
\affiliation{Department of Physics and Astrophysics, University of North Dakota, Grand Forks, ND 58202, USA}
\author{Thomas P. Devereaux}
\affiliation{Stanford Institute for Materials and Energy Sciences, SLAC National Accelerator Laboratory and Stanford University, 2575 Sand Hill Road, Menlo Park, CA 94025, USA}
\affiliation{Department of Materials Science and Engineering, Stanford University, Stanford, CA 94305, USA}
\affiliation{Geballe Laboratory for Advanced Materials, Stanford University, Stanford, CA 94305, USA}

\date{\today}

\begin{abstract}
The bilayer Hubbard model with electron-hole doping is an ideal platform to study excitonic orders due to suppressed recombination via spatial separation of electrons and holes. However, suffering from the sign problem, previous quantum Monte Carlo studies could not arrive at an unequivocal conclusion regarding the presence of phases with clear signatures of excitonic condensation in bilayer Hubbard models. Here, we develop a determinant quantum Monte Carlo (DQMC) algorithm for the bilayer Hubbard model that is sign-problem-free for equal and opposite doping in the two layers, and study excitonic order and charge and spin density modulations as a function of chemical potential difference between the two layers, on-site Coulomb repulsion, and inter-layer interaction. In the intermediate coupling regime and in proximity to the SU(4)-symmetric point, we find a biexcitonic condensate phase at finite electron-hole doping, as well as a competing $(\pi,\pi)$ charge density wave (CDW) state. We extract the Berezinskii-Kosterlitz-Thouless (BKT) transition temperature from superfluid density and a finite size scaling analysis of the correlation functions, and explain our results in terms of an effective biexcitonic hardcore boson model.
\end{abstract}


\maketitle

\textbf{Introduction:} Soon after the prediction of Bose-Einstein condensation in 1926, it was realized that the concept of condensation can be generalized to arbitrary systems of bosonic quasiparticles. One  paradigmatic example has been the BCS ground state of Cooper pairs in superconductors\cite{PhysRev.126.1691, keldysh1968collective}. Exciton condensation is a closely related phenomenon, in which pairs of electrons and holes condense to form a charge-neutral superfluid, and can be understood in a similar manner by invoking an electron-hole transformation. Experimental realizations of exciton condensation have been achieved in quantum Hall bilayers\cite{eisenstein2004bose}, semiconductor quantum wells\cite{high2012condensation}, double bilayer graphene\cite{li2017excitonic}, and in recent experiments demonstrating compelling signatures for exciton condensation in TiSe$_{2}$\cite{Kogar1314} and Ta$_{2}$NiSe$_{5}$\cite{Werdehauseneaap8652}. 
 
A major obstacle for studies of excitonic phenomena has been their short lifetime. Specifically for excitonic ordering, spatial separation of electrons and holes into two layers was proposed to suppress electron-hole recombination\cite{lozovik1976new}. Following this idea, excitonic order, including exciton condensation and biexciton formation, has been extensively studied in electron-hole bilayer continuous models, which describe systems with electrons and holes confined in quantum wells separated into two layers by a barrier\cite{doi:10.1063/1.5052674, PhysRevLett.88.206401, PhysRevLett.110.216407, PhysRevB.78.045313, PhysRevB.79.125308, sharma2018phase}. As well, studies of superconductivity in strongly correlated materials recently have motivated exploration of the electron-hole counterpart in two-band Hubbard-like lattice models\cite{0953-8984-27-33-333201, PhysRevB.88.235115, PhysRevB.88.235127, PhysRevB.85.165135, PhysRevB.90.235140, fujiuchi2018excitonic, PhysRevB.77.144527, PhysRevB.91.144510, PhysRevB.88.035312, 1742-6596-529-1-012030}. However, due to a severe fermion sign problem, a previous DQMC study\cite{PhysRevB.88.235115} of the spin-1/2 bilayer Hubbard model could not arrive at an unequivocal conclusion regarding the existence of exciton condensation. To address this issue, here, we develop a sign-problem-free quantum Monte Carlo algorithm for the bilayer Hubbard model at overall half filling, with equal and opposite doping in each layer, and investigate signatures of excitonic ordering.   

\begin{figure}
\includegraphics[width=\linewidth]{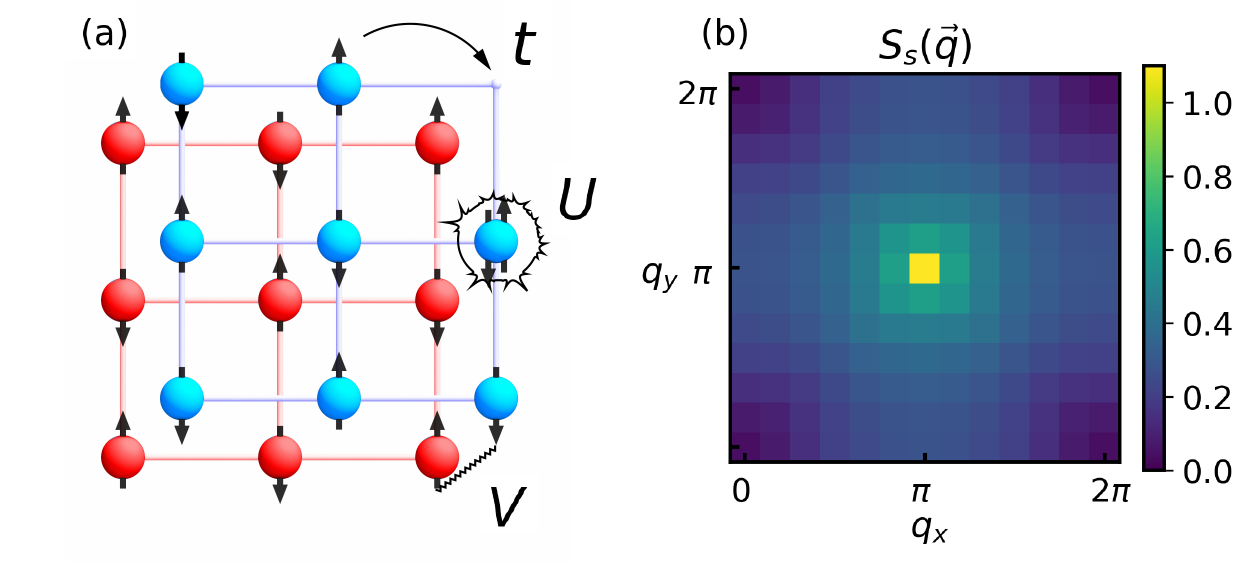}
\caption{\label{fig1}(a) Illustration of parameters in the bilayer Hubbard Hamiltonian Eq.~(\ref{hamiltonian}). (b) Density plot of the spin correlation function $S_{s}(\vec{q})$ in the first Brillouin zone for $U=V=5t$ and $\mu=0t$, a SU(4)-symmetric point. The system size is $L=12$. The inverse temperature is set at $\beta=12/t$.} 
\end{figure}

\textbf{Methods:} The bilayer Hubbard model studied in this work has a form
\begin{align}
\hat{H}&= \hat{H}_{0}+\hat{V}, \nonumber \\
\hat{H_{0}}&=-t\sum\limits_{\langle i,j\rangle,\alpha,\sigma}(\hat{c}^{\dagger}_{i\alpha\sigma}\hat{c}_{j\alpha\sigma}+h.c.)-\mu\sum\limits_{i\sigma}(\hat{n}_{iA\sigma}-\hat{n}_{iB\sigma}), \nonumber \\
\hat{V}&=U\sum\limits_{i\alpha}(\hat{n}_{i\alpha\uparrow}-\frac{1}{2})(\hat{n}_{i\alpha\downarrow}-\frac{1}{2}) \nonumber \\
\ \ \ &+V\sum\limits_{i\sigma\sigma'}(\hat{n}_{iA\sigma}-\frac{1}{2})(\hat{n}_{iB\sigma'}-\frac{1}{2}),
\label{hamiltonian}
\end{align}
split between the kinetic, $\hat{H}_{0}$, and interaction, $\hat{V}$, terms, where $\hat{c}^{\dagger}_{i\alpha\sigma}(\hat{c}_{i\alpha\sigma})$ are creation(annihilation) operators for an electron at site $i$ in layer $\alpha\in \{A, B\}$ with spin $\sigma\in \{\uparrow, \downarrow\}$ and the number operator $\hat{n}_{i\alpha\sigma}\equiv\hat{c}^{\dagger}_{i\alpha\sigma}\hat{c}_{i\alpha\sigma}$. The parameters $t$ and $\mu$ denote the intra-layer hopping amplitude between nearest neighbors and the electron-hole chemical potential, respectively, while $U$ and $V$ are the on-site and inter-layer Coulomb repulsion, respectively. To maintain overall particle-hole symmetry, required by our sign-problem free algorithm, layer A and layer B have equal but opposite filling. All parameters are illustrated schematically in Fig.~\ref{fig1}(a).

We characterize the bilayer Hubbard Hamiltonian in Eq.~(\ref{hamiltonian}) using determinant quantum Monte Carlo (DQMC), a numerically-exact method to simulate interacting quantum many-body systems at finite temperature. Detailed introductions to the DQMC algorithm can be found in Ref.~\cite{PhysRevD.24.2278,PhysRevB.40.506,santos2003introduction}. 

In general, the Hubbard-Stratonovich (HS) field configuration dependent Boltzmann weight is
\begin{align}
    w_{\mathbf{s}}=\det \left[ \mathbf{I}+\mathbf{B}_{\mathbf{s}}\right],
    \label{det}
\end{align}
where $\mathbf{I}$ is the identity matrix and $\mathbf{B}_{\mathbf{s}}$ is a matrix depends on the the configuration $\mathbf{s}$. The determinant per HS field configuration in Eq.~(\ref{det}) can be negative (or even complex) for fermions in generic cases, which is known as the fermion sign problem, giving rise to large statistical errors and restricting simulations to relatively high temperatures.

\begin{figure*}
\includegraphics[width=0.95\textwidth]{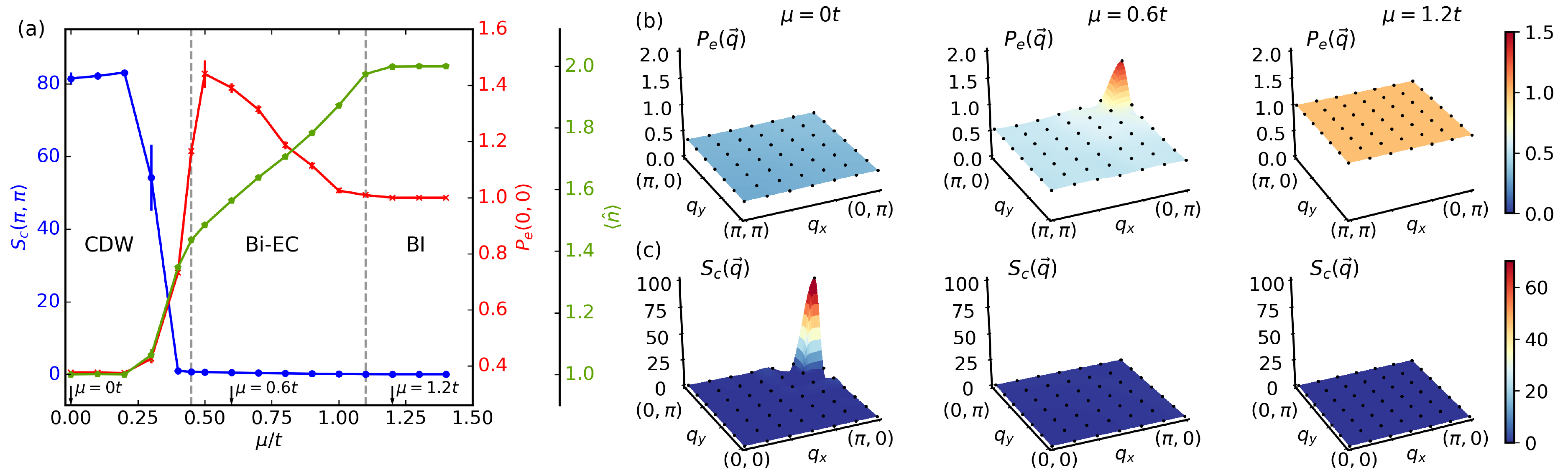}
\caption{\label{fig2} Summary of results for $U=5t$, $V=6t$ and electron-hole chemical potentials ranging from $\mu=0t$ to $\mu=1.4t$, obtained with a system of linear size $L=12$ at inverse temperature $\beta=12/t$. (a) $S_{c}(\pi,\pi)$ (blue), $P_{e}(0,0)$ (red) and $\langle\hat{n}\rangle$ (green) plotted as functions of $\mu$, with error bars for each data point taken from Monte Carlo estimates. The error bars are of the size of the marker for most of the data points. Grey dashed lines separating CDW, Bi-EC, and BI phases are guides to the eye. (b) $P_{e}(\vec{q})$ in the first Brillouin zone at $\mu=0t$ (left), $\mu=0.6t$ (middle) and $\mu=1.2t$ (right). (c) $S_{c}(\vec{q})$ in the first Brillouin zone at $\mu=0t$ (left), $\mu=0.6t$ (middle) and $\mu=1.2t$ (right). The black points in (b) and (c) identify actual data points from the calculation, whereas the colored surfaces are interpolated from these data points.}
\end{figure*} 

Over the past few years, a number of algorithms were proposed to ``solve" this fermion sign problem for specific models and parameter regimes\cite{koonin1997shell, PhysRevLett.83.3116, imada2000path, PhysRevB.71.155115, berg2012sign, PhysRevLett.111.130402, li2015solving, PhysRevLett.116.250601}. One possible strategy to prove that the probability weights are positive semi-definite is to show that $\mathbf{I}+\mathbf{B}_{\mathbf{s}}$ has an anti-unitary symmetry $\mathbf{T}$, i.e. $\mathbf{T}^2 = -\mathbf{I}$ and $\mathbf{T}^{-1}(\mathbf{I}+\mathbf{B}_{\mathbf{s}})\mathbf{T}=\mathbf{I}+\mathbf{B}_{\mathbf{s}}$\cite{koonin1997shell}. To achieve this, we first perform a single-layer particle-hole transformation on the Hamiltonian $\hat{H}$: $\hat{c}_{iB\sigma}\rightarrow(-1)^{\delta_{i}}\hat{c}^{\dagger}_{iB\sigma}$, where $\delta_{i}$ is even/odd on neighboring site. Then the interacting part of the Hamiltonian $\hat{V}$ can be decomposed by introducing two sets of spin-1 HS field configurations $\{\mathbf{s}\}$ and $\{\bar{\mathbf{s}}\}$ in spacetime, each taking values $\{-1, 0, 1\}$. Consider now an unconventional anti-unitary symmetry
\begin{align}
\hat{T} &= \sum_{i\sigma}[|i,A,\sigma\rangle\langle i,B,\sigma |-|i,B,\sigma\rangle\langle i,A,\sigma |]\hat{K},
\end{align}
where $\hat{K}$ is the complex-conjugation operator. Matrix $\mathbf{T}$ is related to operator $\hat{T}$ by $\mathbf{c}^{\dagger}\mathbf{T}\mathbf{c}=\hat{T}$, where $\mathbf{c}=(..., c_{i,A,\uparrow}, c_{i,A,\downarrow}, c_{i,B,\uparrow}, c_{i,B,\downarrow}, ...)$. One finds that $\mathbf{I}+\mathbf{B}_{\mathbf{s}}$ is symmetric under $\mathbf{T}$ when
\begin{eqnarray}
|U|\leq V,
\label{signfree}
\end{eqnarray} 
which in turn determines the sign-problem-free parameter regime of the algorithm. 
 
Details for the sign-problem-free algorithm are given in the Supplemental Material. The strategy proposed here works for all bipartite lattices, including square and honeycomb lattices. Here, we focus on presenting results for the square lattice with periodic boundary condition. 

\textbf{Results:} 
We start by presenting results at $U=V$ and $\mu=0t$, where the Hamiltonian is SU(4) symmetric\cite{PhysRevB.61.12112, PhysRevLett.93.016407}. Charge and spin modulations are described by the charge correlation function $S_{c}(\vec{q})$ and the spin correlation function $S_{s}(\vec{q})$ defined in the usual way (see the Supplemental Material for definitions). Particle-hole symmetry between the two layers ensures that we can restrict measurements of the charge and spin correlation functions to the electron-doped layer (layer A). Fig.~\ref{fig1}(b) shows the spin correlation function $S_{s}(\vec{q})$ for $U=V=5t$ and $\mu=0$, measured with a system of linear size $L=12$ at inverse temperature $\beta=12/t$. There is a clear enhancement of $S_{s}(\vec{q})$ at $\vec{q}=(\pi,\pi)$, while the charge correlation function $S_{c}(\vec{q})$ shows no feature here (see the Supplemental Material), indicating SU(4) antiferromagnetism.

Exciton condensation may be obtained by breaking the U(1) symmetry for the conserved charge $\sum_{i \sigma} \hat{n}_{i A \sigma} - \hat{n}_{i B \sigma}$. However, with the SU(4) symmetry, the invariance of $\hat{H}$ under transformation $\ A\downarrow \leftrightarrow B\uparrow$ entails the equivalence between excitonic and spin ordering, which is confirmed numerically as shown in the Supplementary Material. It is therefore natural to inquire about instabilities towards excitonic condensation upon a weak breaking of the SU(4) symmetry. In the following, we hence choose $U=5t, V=6t$, which reduces SU(4) to SU(2)$\times$SU(2), representing two independent spin rotational symmetries for layer A and layer B, to lift the degeneracy between excitonic and spin ordering, and systematically study the phase diagram as a function of varying electron-hole doping $\mu$ in the two layers.

Consider now the order parameter for a conventional excitonic condensate in either singlet or triplet channels, described generically by single-exciton creation operators of the form $\hat{b}^{\dagger}_{\vec{q}} = \sum_{\vec{k}\sigma\sigma'} \hat{c}^{\dagger}_{\vec{k}+\vec{q}A\sigma} \tau^{\sigma\sigma'}  \hat{c}_{\vec{k}B\sigma'}$. Such a phase breaks both the excitonic U(1) and residual SU(2)$\times$SU(2) spin rotation symmetry, precluding a finite-temperature phase transition by the Mermin-Wagner theorem\cite{PhysRevLett.17.1133}. Instead, intriguingly, we can define a biexciton creation/annihilation operator
\begin{align}
    \hat{\Delta}(\vec{r})=\hat{c}^{\dagger}_{\vec{r}B\uparrow}\hat{c}^{\dagger}_{\vec{r}B\downarrow}\hat{c}_{\vec{r}A\downarrow}\hat{c}_{\vec{r}A\uparrow}
    \label{pairfield}
\end{align}
which breaks the excitonic U(1) symmetry but preserves SU(2)$\times$SU(2). It is therefore possible to obtain a BKT transition \cite{berezinskii1972overview, Kosterlitz_1973, kosterlitz1974critical} to a quasi-long-range biexciton condensate at finite temperature. We therefore focus on studying the corresponding correlation function
\begin{align}
    P_{e}(\vec{q}) = \frac{1}{L^{2}}\sum_{\vec{R},\vec{r}}e^{-i\vec{q}\cdot\vec{R}}\langle\hat{\Delta}^{\dagger}(\vec{R}+\vec{r})\hat{\Delta}(\vec{r})\rangle.
\end{align} 

\begin{figure}
\includegraphics[width=0.9\linewidth]{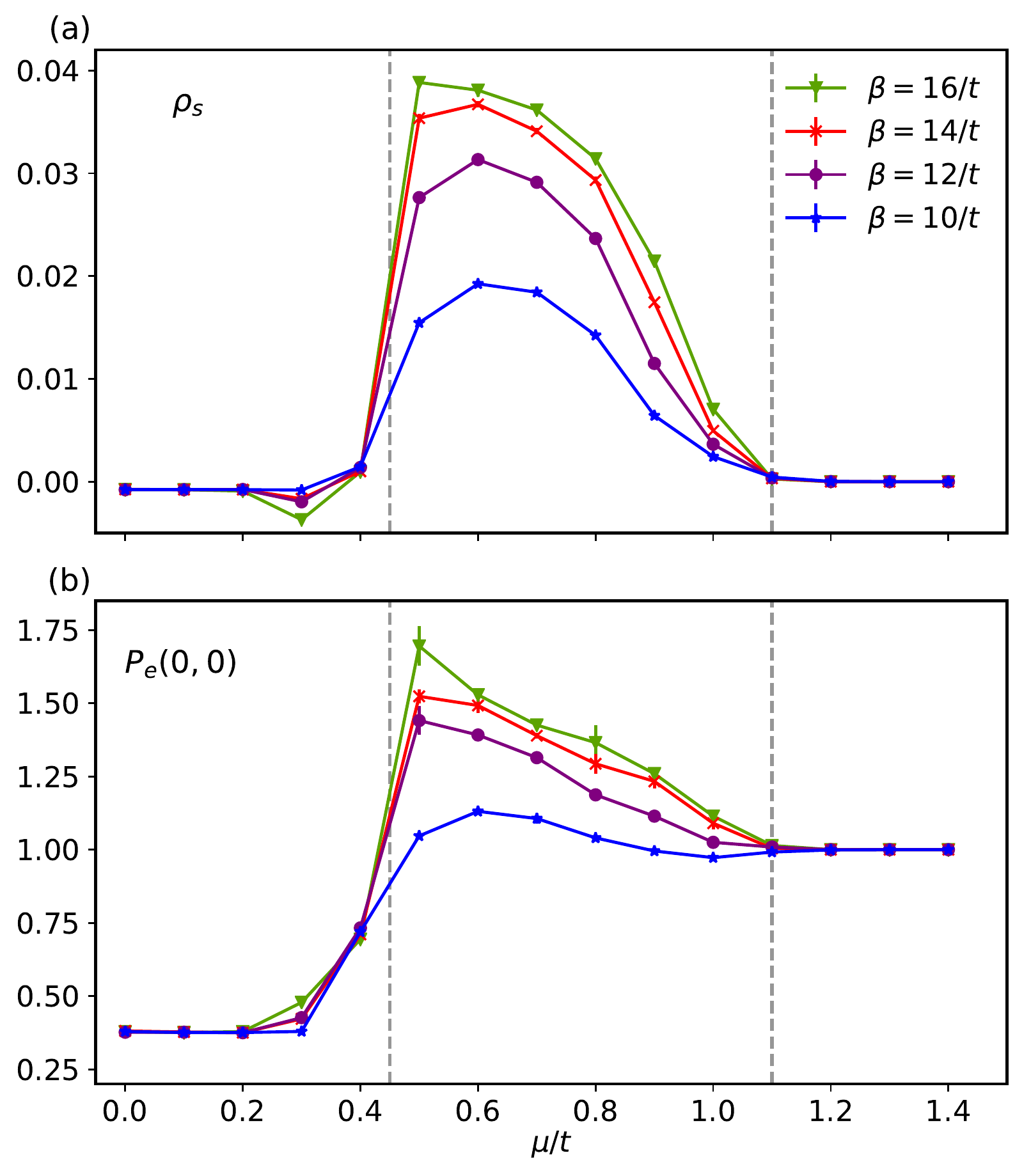}
\caption{\label{fig3} (a) $\rho_{s}$ calculated for $U=5t$, $V=6t$ and electron-hole chemical potentials ranging from $\mu=0t$ to $\mu=1.4t$ for a system of size $L=12$. (b) $P_{e}(0,0)$ for the same parameters. The grey dashed lines are guides to the eye, indicating the same chemical potentials separating the phases as in Fig.~\ref{fig2}(a).}
\end{figure} 

\begin{figure}
\includegraphics[width=0.9\linewidth]{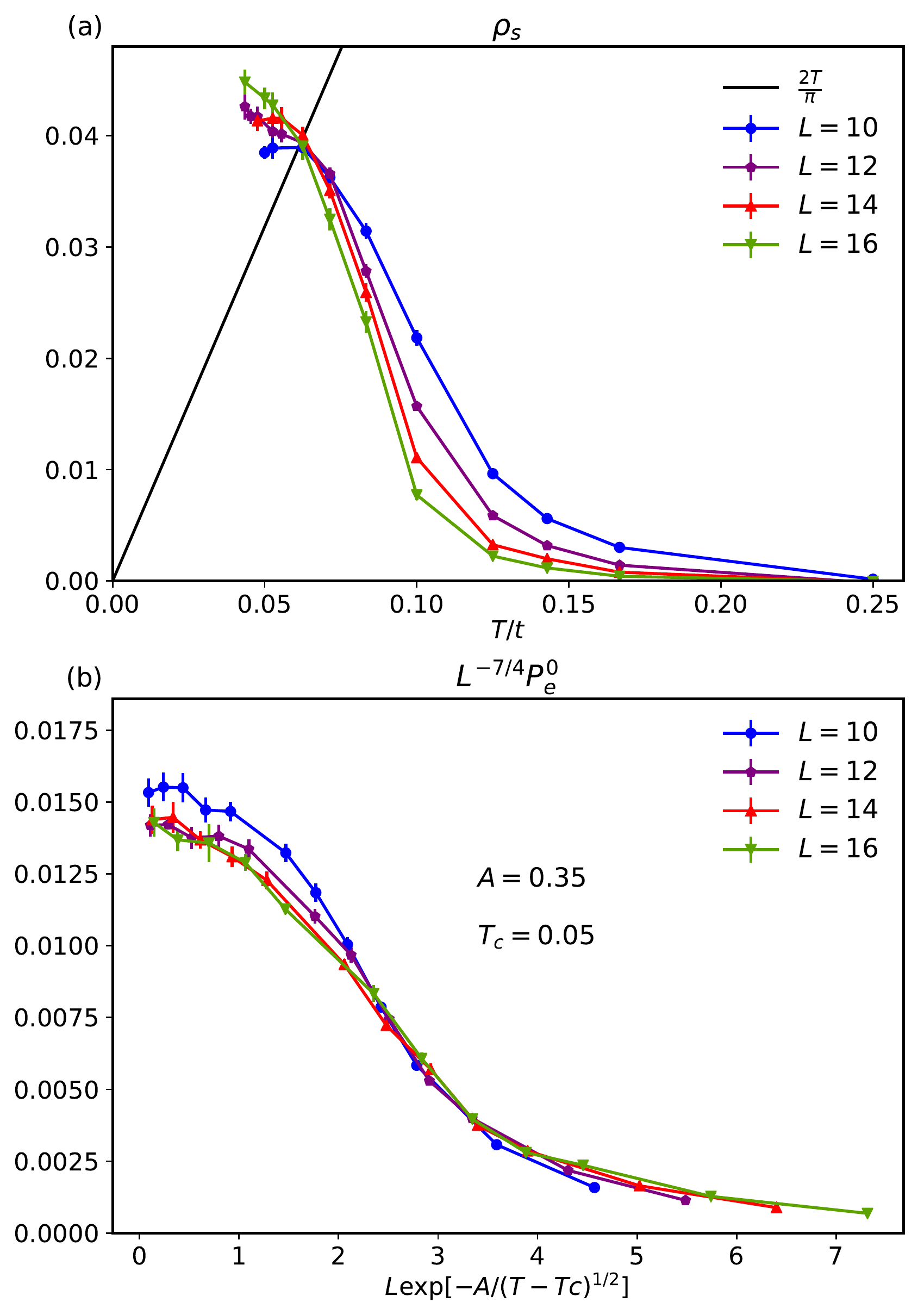}
\caption{\label{fig4} (a) SF density $\rho_{s}$ calculated for $U=5t$, $V=6t$ and $\mu=0.5t$ with system sizes from $L=10$ to $L=16$. The black solid line is $\rho=2T/\pi$. The intercept of the black solid line and SF density data points determines the BKT transition temperature $T_{c}$. (b) Plot of normalized correlation $L^{-7/4}P^{0}_{e}$ versus $L/\xi=L\exp[-A/(T-T_{c})^{1/2}]$ for $A=0.35$ and $T_{c}=0.05$ (units are omitted). A good data collapse is especially achieved for $L\geq 12$. The unscaled plot of $L^{-7/4}P^{0}_{e}$ as a function of $\beta$ can be found in the Supplemental Material.}
\end{figure} 

Fig.~\ref{fig2} summarizes results of various ordering tendencies for $U=5t$, $V=6t$ and electron-hole chemical potentials ranging from $\mu=0t$ to $\mu=1.4t$, obtained with a system of linear size $L=12$ at inverse temperature $\beta=12/t$. At half-filling, the charge correlation function $S
_{c}(\vec{q})$ shows a peak at $\vec{q}=(\pi,\pi)$, which implies CDW order. As $\mu$ increases and the two layers become electron-hole doped, the $(\pi,\pi)$ charge correlation peak is suppressed, accompanied by a gradual increase in biexciton correlations at $\vec{q}=(0,0)$, indicating biexciton condensation (Bi-EC). The CDW and Bi-EC appear to compete. Upon further increase of $\mu$, the electron doped layer eventually becomes fully filled and the system enters a band insulator (BI) phase. Spin orders are absent throughout the electron-hole chemical potential range we have studied. Fig.~\ref{fig2}(a) shows $S_{c}(\pi,\pi)$, $P_{e}(0,0)$ and $\langle\hat{n}\rangle$, the average electron density per site in the electron doped layer, as functions of $\mu$. For three typical $\mu$ values: $0t$, $0.6t$ and $1.2t$, density plots of $P_{e}$ and $S_{c}$ in momentum space are presented in Fig.~\ref{fig2}(b) and Fig.~\ref{fig2}(c), respectively. At $\mu=0t$, the $(\pi,\pi)$ CDW order gives rise to enhanced local biexcitonic binding in real space, which results in a finite and uniform distribution of $P_{e}(\vec{q})$ in momentum space. At $\mu=0.6t$, the CDW order completely disappears and a biexciton correlation peak emerges at $\vec{q}=(0,0)$. Finally, $\mu=1.2t$ falls in the BI regime where $\langle\hat{n}\rangle=2$, with all ordering tendencies absent.    

To confirm the existence of biexciton condensation and extract the BKT transition temperature $T_{c}$, we have examined the superfluid (SF) density $\rho_{s}$ and performed a finite size scaling analysis on $P_{e}$. Before proceeding, we note that Bi-EC and SF refer to the same state of matter here, with the U(1) symmetry of the phase factor $\theta$ of $\hat{\Delta}$ broken. $\rho_{s}$ is defined by the classical action for phase fluctuations:
\begin{align}
    S=\frac{\beta}{2}\int d^{2}r\rho_{s}(\nabla\theta)^{2},
\end{align}
and is related to unequal-time current-current correlation functions by \cite{scalapino1993insulator, xu2017non}
\begin{align}
    \rho_{s}=&\frac{1}{16}(\delta\Lambda^{xx}_{AA}+\delta\Lambda^{xx}_{BB}-2\delta\Lambda^{xx}_{AB}), \\
    \delta\Lambda^{xx}_{\alpha\gamma}=&\lim_{q_{x}\rightarrow 0}\Lambda^{xx}_{\alpha\gamma}(q_{x},q_{y}=0,\omega_{n}=0) \nonumber \\
    -&\lim_{q_{y}\rightarrow 0}\Lambda^{xx}_{\alpha\gamma}(q_{x}=0,q_{y},\omega_{n}=0),
\end{align}
where
\begin{align}
    \Lambda^{xx}_{\alpha\gamma}(\vec{q},\omega_{n})=&\sum_{\vec{r}}\int^{\beta}_{0}d\tau e^{-i\vec{q}\cdot\vec{r}}e^{i\omega_{n}\tau}\Lambda^{xx}_{\alpha\gamma}(\vec{r},\tau), \\
    \Lambda^{xx}_{\alpha\gamma}(\vec{r},\tau)&=\sum_{\sigma,\sigma'}\langle \hat{j}^{x}_{\alpha,\sigma}(\vec{r},\tau)\hat{j}^{x}_{\gamma,\sigma'}(0,0)\rangle.
\end{align}
Here, $\hat{j}^{x}_{\alpha\sigma}(\vec{r},\tau)$ is the x component of the current operator for imaginary time $\tau$ (see the Supplemental Material).

Fig.~\ref{fig3} compares $\rho_{s}$ and $P_{e}(0,0)$ as a function of $\mu$ and temperature. For $\mu$ small (CDW) or large (BI), $\rho_{s}$ is always zero and independent of temperature. In contrast, in the Bi-EC (SF) region, $\rho_{s}$ becomes finite and increases upon increasing the inverse temperature $\beta$, which signals a transition of the system to a Bi-EC phase. The slight dips of $\rho_{s}$ before the onset of Bi-EC are finite size artifacts. Similar behaviors are found also in $P_{e}(0,0)$, as shown in Fig.~\ref{fig3}(b). The consistency between $\rho_{s}$ and $P_{e}(0,0)$ establishes the reliability of the SF density calculation and further corroborates the existence of the Bi-EC phase.

Next, we proceed to a finite size analysis and $T_{c}$ extrapolation, following the method used in a previous work which systematically determines the critical temperature of superconductivity in the 2D attractive Hubbard model\cite{PhysRevB.69.184501}. In numerical studies, a universal-jump of the SF density is considered as a signature of a BKT transition. Approaching $T_{c}$ from below, the following relation is satisfied \cite{PhysRevLett.39.1201}:
\begin{align}
    T_{c}=\frac{\pi}{2}\rho_{s}.
    \label{tc}
\end{align}
Fig.~\ref{fig4}(a) displays $\rho_{s}$ measured for $U=5t$, $V=6t$ and $\mu=0.5t$, with linear system sizes up to $L=16$ at temperatures down to $T=t/24$. A clear jump of the value of $\rho_{s}$ shows up for every system size upon decreasing temperature and becomes more abrupt for larger system sizes. $\rho=2T/\pi$ is plotted with a black solid line in Fig.~\ref{fig4}(a); the intercept of $\rho=2T/\pi$ and $\rho_{s}$ gives an estimation for $T_{c}\approx0.06t$.

Another approach to extract $T_{c}$ is by finite size scaling analysis of $P_{e}(0,0)$. For $T\leq T_{c}$, the correlation strength falls algebraically as $\langle\hat{\Delta}^{\dagger}(\vec{r})\hat{\Delta}(0)\rangle\sim |\vec{r}|^{-\eta(T)}$, with $\eta(T_{c})=0.25$ at the transition temperature \cite{Kosterlitz_1973, berche2002correlations}. For finite systems with $L\gg 1$ and $T\rightarrow T_{c}$ from above, $P_{e}(0,0)$ follows the relation\cite{PhysRevLett.66.946}
\begin{align}
    P_{e}&(0,0)=L^{2-\eta(T_{c})}f(L/\xi), \label{fs}\\
    &\xi\sim\exp\left[\frac{A}{(T-T_{c})^{1/2}}\right].
\end{align}
$T_{c}$ is determined by adjusting $A$ and $T_{c}$ until one reaches optimal data collapse. For a faster convergence with respect to system size, instead of $P_{e}(0,0)$ we perform the analysis on the local correlation subtracted value $P^{0}_{e}=P_{e}(0,0)-\frac{1}{L^{2}}\sum_{\vec{r}}\langle\hat{\Delta}^{\dagger}(\vec{r})\hat{\Delta}(\vec{r})\rangle$, which should follow the same relation as $P_{e}(0,0)$ [Eq.~(\ref{fs})]. As presented in Fig.~\ref{fig4}(b), a good data collapse can be achieved with $A=0.35$ and $T_{c}=0.05t$. The unscaled plot of $L^{-7/4}P^{0}_{e}$ can be found in the Supplemental Material. We find that the $T_{c}$ estimated by two independent approaches are roughly consistent with each other and serve as convincing evidence for a BKT transition to a Bi-EC (SF) phase.

\textbf{Discussion:} In the strong coupling regime where $U,V\gg t$ and $V-U\gg t$, the low-energy physics of the bilayer Hubbard Hamiltonian $\hat{H}$ can be captured by an effective hard-core boson model
\begin{align}
    \hat{H}_{\rm eff} = -t_b \sum_{\left<ij\right>} \hat{a}_i^{\dag} \hat{a}_j^{\vphantom{\dag}} + V_b \sum_{\left<ij\right>} \hat{n}_i^a \hat{n}_j^a - \mu_b \sum_i \hat{n}_i^a
\end{align}
that constrains the local configuration of each site to either two electrons in layer A or B, neglecting charge fluctuations at finite $V$. In this basis, spanned by hardcore boson operators $\hat{a}_{i} \equiv \hat{\Delta}(\vec{r}_i)$ [Eq. (\ref{pairfield})], virtual hopping of fermions perturbatively generates nearest-neighbor boson repulsion $V_b =\frac{8t^2}{2V-U}+\frac{4t^4}{(2V-U)^2} \left[ \frac{4}{V-U} + \frac{1}{V} - \frac{8}{2V-U} \right]$ and boson hopping $t_b = \frac{2 t^4}{(2V-U)^2} \left[ \frac{4}{V-U} + \frac{1}{V} \right]$ respectively. Finally, the electron-hole doping $\mu$ enters equivalently as a chemical potential for the hardcore bosons with $\mu_b = 4\mu - 2V_b$. Refer to the Supplemental Material for details.

Competition between the CDW and Bi-EC orders can thus be understood in terms of the well-known competition between checkerboard order and superfluid phase in the hardcore boson model \cite{PhysRevB.56.3212, PhysRevLett.84.1599, PhysRevLett.88.167208}. Importantly, this picture indicates the stability of the Bi-EC in a wide parameter range at zero temperature. However, we emphasize that studying the high transition temperature regime at intermediate coupling strength necessitates accounting for significant charge fluctuations, beyond the boson mapping, making a rigorous numerical analysis indispensable.

Finally, it also is noteworthy that after performing a particle-hole transformation on layer B, the electron-hole bilayer Hubbard model maps onto an electron-doped bilayer Hubbard model with attractive inter-layer interaction and repulsive on-site Coulomb interaction. In this scenario, Bi-EC order corresponds to exotic charge-4e superconductivity (SC). While exciton condensation and charge-2e SC have been scrutinized both theoretically and experimentally, Bi-EC and charge-4e SC remain relatively under-explored, mainly due to the four-particle nature of their order parameters, and are energetically less favorable than their two-particle counterparts in most cases. Recent developments include identification of Bi-EC in a two-orbital $t-J$ chain\cite{PhysRevB.98.035128} and exploration of charge-4e SC using Majorana quantum Monte Carlo\cite{PhysRevB.95.241103}.

\textbf{Conclusion:} In summary we have developed a sign-problem free DQMC algorithm for the bilayer Hubbard model in the parameter regime $|U| \leq V$. With this tool, we examined competition between spin, charge, and excitonic orders on the square lattice. Remarkably, away from the SU(4) point, we find convincing numerical evidence for a biexcitonic condensate, which competes with $(\pi,\pi)$ charge order at finite electron-hole doping. We have extracted the BKT transition temperature from the superfluid density, as well as a finite size and temperature scaling analysis of biexciton correlation functions, with an estimate for $T_{c} \sim 0.05t-0.06t$. Finally, our algorithm applies to any bipartite lattice including the honeycomb lattice, and can be straightforwardly extended to include magnetic fields and inter-layer hopping. Thus, this algorithm can be extended to study the interplay of competing orders in a variety of systems such as bilayer graphene. We expect that such exact numerical studies of electron-hole condensates will advance our understanding of quasiparticle condensation in general, and ultimately may shed light on the strong correlation driven mechanism behind Cooper pair condensation in unconventional superconductors.

\begin{acknowledgments}
The authors thank Y. Schattner and J. Zaanen for illuminating discussions. X.X.H., E.W.H., B.M. and T.P.D. acknowledge support from the US Department of Energy, Office of Science, Office of Basic Energy Sciences, Division of Materials Sciences and Engineering, under Contract No. DE-AC02-76SF00515. M.C. is supported by the Flatiron Institute, a division of the Simons Foundation. The authors thank Sirui Chen for designing Fig.~\ref{fig1}(a).
\end{acknowledgments}

\bibliography{cite}

\begin{thebibliography}{52}%
\makeatletter
\providecommand \@ifxundefined [1]{%
 \@ifx{#1\undefined}
}%
\providecommand \@ifnum [1]{%
 \ifnum #1\expandafter \@firstoftwo
 \else \expandafter \@secondoftwo
 \fi
}%
\providecommand \@ifx [1]{%
 \ifx #1\expandafter \@firstoftwo
 \else \expandafter \@secondoftwo
 \fi
}%
\providecommand \natexlab [1]{#1}%
\providecommand \enquote  [1]{``#1''}%
\providecommand \bibnamefont  [1]{#1}%
\providecommand \bibfnamefont [1]{#1}%
\providecommand \citenamefont [1]{#1}%
\providecommand \href@noop [0]{\@secondoftwo}%
\providecommand \href [0]{\begingroup \@sanitize@url \@href}%
\providecommand \@href[1]{\@@startlink{#1}\@@href}%
\providecommand \@@href[1]{\endgroup#1\@@endlink}%
\providecommand \@sanitize@url [0]{\catcode `\\12\catcode `\$12\catcode
  `\&12\catcode `\#12\catcode `\^12\catcode `\_12\catcode `\%12\relax}%
\providecommand \@@startlink[1]{}%
\providecommand \@@endlink[0]{}%
\providecommand \url  [0]{\begingroup\@sanitize@url \@url }%
\providecommand \@url [1]{\endgroup\@href {#1}{\urlprefix }}%
\providecommand \urlprefix  [0]{URL }%
\providecommand \Eprint [0]{\href }%
\providecommand \doibase [0]{http://dx.doi.org/}%
\providecommand \selectlanguage [0]{\@gobble}%
\providecommand \bibinfo  [0]{\@secondoftwo}%
\providecommand \bibfield  [0]{\@secondoftwo}%
\providecommand \translation [1]{[#1]}%
\providecommand \BibitemOpen [0]{}%
\providecommand \bibitemStop [0]{}%
\providecommand \bibitemNoStop [0]{.\EOS\space}%
\providecommand \EOS [0]{\spacefactor3000\relax}%
\providecommand \BibitemShut  [1]{\csname bibitem#1\endcsname}%
\let\auto@bib@innerbib\@empty
\bibitem [{\citenamefont {Blatt}\ \emph {et~al.}(1962)\citenamefont {Blatt},
  \citenamefont {B\"oer},\ and\ \citenamefont {Brandt}}]{PhysRev.126.1691}%
  \BibitemOpen
  \bibfield  {author} {\bibinfo {author} {\bibfnamefont {J.~M.}\ \bibnamefont
  {Blatt}}, \bibinfo {author} {\bibfnamefont {K.~W.}\ \bibnamefont {B\"oer}}, \
  and\ \bibinfo {author} {\bibfnamefont {W.}~\bibnamefont {Brandt}},\ }\href
  {\doibase 10.1103/PhysRev.126.1691} {\bibfield  {journal} {\bibinfo
  {journal} {Phys. Rev.}\ }\textbf {\bibinfo {volume} {126}},\ \bibinfo {pages}
  {1691} (\bibinfo {year} {1962})}\BibitemShut {NoStop}%
\bibitem [{\citenamefont {Keldysh}\ and\ \citenamefont
  {Kozlov}(1968)}]{keldysh1968collective}%
  \BibitemOpen
  \bibfield  {author} {\bibinfo {author} {\bibfnamefont {L.}~\bibnamefont
  {Keldysh}}\ and\ \bibinfo {author} {\bibfnamefont {A.}~\bibnamefont
  {Kozlov}},\ }\href@noop {} {\bibfield  {journal} {\bibinfo  {journal} {Sov.
  Phys. JETP}\ }\textbf {\bibinfo {volume} {27}},\ \bibinfo {pages} {521}
  (\bibinfo {year} {1968})}\BibitemShut {NoStop}%
\bibitem [{\citenamefont {Eisenstein}\ and\ \citenamefont
  {MacDonald}(2004)}]{eisenstein2004bose}%
  \BibitemOpen
  \bibfield  {author} {\bibinfo {author} {\bibfnamefont {J.}~\bibnamefont
  {Eisenstein}}\ and\ \bibinfo {author} {\bibfnamefont {A.}~\bibnamefont
  {MacDonald}},\ }\href@noop {} {\bibfield  {journal} {\bibinfo  {journal}
  {Nature}\ }\textbf {\bibinfo {volume} {432}},\ \bibinfo {pages} {691}
  (\bibinfo {year} {2004})}\BibitemShut {NoStop}%
\bibitem [{\citenamefont {High}\ \emph {et~al.}(2012)\citenamefont {High},
  \citenamefont {Leonard}, \citenamefont {Remeika}, \citenamefont {Butov},
  \citenamefont {Hanson},\ and\ \citenamefont
  {Gossard}}]{high2012condensation}%
  \BibitemOpen
  \bibfield  {author} {\bibinfo {author} {\bibfnamefont {A.}~\bibnamefont
  {High}}, \bibinfo {author} {\bibfnamefont {J.}~\bibnamefont {Leonard}},
  \bibinfo {author} {\bibfnamefont {M.}~\bibnamefont {Remeika}}, \bibinfo
  {author} {\bibfnamefont {L.}~\bibnamefont {Butov}}, \bibinfo {author}
  {\bibfnamefont {M.}~\bibnamefont {Hanson}}, \ and\ \bibinfo {author}
  {\bibfnamefont {A.}~\bibnamefont {Gossard}},\ }\href@noop {} {\bibfield
  {journal} {\bibinfo  {journal} {Nano letters}\ }\textbf {\bibinfo {volume}
  {12}},\ \bibinfo {pages} {2605} (\bibinfo {year} {2012})}\BibitemShut
  {NoStop}%
\bibitem [{\citenamefont {Li}\ \emph {et~al.}(2017)\citenamefont {Li},
  \citenamefont {Taniguchi}, \citenamefont {Watanabe}, \citenamefont {Hone},\
  and\ \citenamefont {Dean}}]{li2017excitonic}%
  \BibitemOpen
  \bibfield  {author} {\bibinfo {author} {\bibfnamefont {J.}~\bibnamefont
  {Li}}, \bibinfo {author} {\bibfnamefont {T.}~\bibnamefont {Taniguchi}},
  \bibinfo {author} {\bibfnamefont {K.}~\bibnamefont {Watanabe}}, \bibinfo
  {author} {\bibfnamefont {J.}~\bibnamefont {Hone}}, \ and\ \bibinfo {author}
  {\bibfnamefont {C.}~\bibnamefont {Dean}},\ }\href@noop {} {\bibfield
  {journal} {\bibinfo  {journal} {Nature Physics}\ }\textbf {\bibinfo {volume}
  {13}},\ \bibinfo {pages} {751} (\bibinfo {year} {2017})}\BibitemShut
  {NoStop}%
\bibitem [{\citenamefont {Kogar}\ \emph {et~al.}(2017)\citenamefont {Kogar},
  \citenamefont {Rak}, \citenamefont {Vig}, \citenamefont {Husain},
  \citenamefont {Flicker}, \citenamefont {Joe}, \citenamefont {Venema},
  \citenamefont {MacDougall}, \citenamefont {Chiang}, \citenamefont {Fradkin},
  \citenamefont {van Wezel},\ and\ \citenamefont {Abbamonte}}]{Kogar1314}%
  \BibitemOpen
  \bibfield  {author} {\bibinfo {author} {\bibfnamefont {A.}~\bibnamefont
  {Kogar}}, \bibinfo {author} {\bibfnamefont {M.~S.}\ \bibnamefont {Rak}},
  \bibinfo {author} {\bibfnamefont {S.}~\bibnamefont {Vig}}, \bibinfo {author}
  {\bibfnamefont {A.~A.}\ \bibnamefont {Husain}}, \bibinfo {author}
  {\bibfnamefont {F.}~\bibnamefont {Flicker}}, \bibinfo {author} {\bibfnamefont
  {Y.~I.}\ \bibnamefont {Joe}}, \bibinfo {author} {\bibfnamefont
  {L.}~\bibnamefont {Venema}}, \bibinfo {author} {\bibfnamefont {G.~J.}\
  \bibnamefont {MacDougall}}, \bibinfo {author} {\bibfnamefont {T.~C.}\
  \bibnamefont {Chiang}}, \bibinfo {author} {\bibfnamefont {E.}~\bibnamefont
  {Fradkin}}, \bibinfo {author} {\bibfnamefont {J.}~\bibnamefont {van Wezel}},
  \ and\ \bibinfo {author} {\bibfnamefont {P.}~\bibnamefont {Abbamonte}},\
  }\href {\doibase 10.1126/science.aam6432} {\bibfield  {journal} {\bibinfo
  {journal} {Science}\ }\textbf {\bibinfo {volume} {358}},\ \bibinfo {pages}
  {1314} (\bibinfo {year} {2017})}\BibitemShut {NoStop}%
\bibitem [{\citenamefont {Werdehausen}\ \emph {et~al.}(2018)\citenamefont
  {Werdehausen}, \citenamefont {Takayama}, \citenamefont {H{\"o}ppner},
  \citenamefont {Albrecht}, \citenamefont {Rost}, \citenamefont {Lu},
  \citenamefont {Manske}, \citenamefont {Takagi},\ and\ \citenamefont
  {Kaiser}}]{Werdehauseneaap8652}%
  \BibitemOpen
  \bibfield  {author} {\bibinfo {author} {\bibfnamefont {D.}~\bibnamefont
  {Werdehausen}}, \bibinfo {author} {\bibfnamefont {T.}~\bibnamefont
  {Takayama}}, \bibinfo {author} {\bibfnamefont {M.}~\bibnamefont
  {H{\"o}ppner}}, \bibinfo {author} {\bibfnamefont {G.}~\bibnamefont
  {Albrecht}}, \bibinfo {author} {\bibfnamefont {A.~W.}\ \bibnamefont {Rost}},
  \bibinfo {author} {\bibfnamefont {Y.}~\bibnamefont {Lu}}, \bibinfo {author}
  {\bibfnamefont {D.}~\bibnamefont {Manske}}, \bibinfo {author} {\bibfnamefont
  {H.}~\bibnamefont {Takagi}}, \ and\ \bibinfo {author} {\bibfnamefont
  {S.}~\bibnamefont {Kaiser}},\ }\href {\doibase 10.1126/sciadv.aap8652}
  {\bibfield  {journal} {\bibinfo  {journal} {Science Advances}\ }\textbf
  {\bibinfo {volume} {4}} (\bibinfo {year} {2018}),\
  10.1126/sciadv.aap8652}\BibitemShut {NoStop}%
\bibitem [{\citenamefont {Lozovik}\ and\ \citenamefont
  {Yudson}(1976)}]{lozovik1976new}%
  \BibitemOpen
  \bibfield  {author} {\bibinfo {author} {\bibfnamefont {Y.~E.}\ \bibnamefont
  {Lozovik}}\ and\ \bibinfo {author} {\bibfnamefont {V.}~\bibnamefont
  {Yudson}},\ }\href@noop {} {\bibfield  {journal} {\bibinfo  {journal} {Zh.
  Eksp. i Teor. Fiz}\ }\textbf {\bibinfo {volume} {71}},\ \bibinfo {pages}
  {738} (\bibinfo {year} {1976})}\BibitemShut {NoStop}%
\bibitem [{\citenamefont {Fil}\ and\ \citenamefont
  {Shevchenko}(2018)}]{doi:10.1063/1.5052674}%
  \BibitemOpen
  \bibfield  {author} {\bibinfo {author} {\bibfnamefont {D.~V.}\ \bibnamefont
  {Fil}}\ and\ \bibinfo {author} {\bibfnamefont {S.~I.}\ \bibnamefont
  {Shevchenko}},\ }\href {\doibase 10.1063/1.5052674} {\bibfield  {journal}
  {\bibinfo  {journal} {Low Temperature Physics}\ }\textbf {\bibinfo {volume}
  {44}},\ \bibinfo {pages} {867} (\bibinfo {year} {2018})},\ \Eprint
  {http://arxiv.org/abs/https://doi.org/10.1063/1.5052674}
  {https://doi.org/10.1063/1.5052674} \BibitemShut {NoStop}%
\bibitem [{\citenamefont {De~Palo}\ \emph {et~al.}(2002)\citenamefont
  {De~Palo}, \citenamefont {Rapisarda},\ and\ \citenamefont
  {Senatore}}]{PhysRevLett.88.206401}%
  \BibitemOpen
  \bibfield  {author} {\bibinfo {author} {\bibfnamefont {S.}~\bibnamefont
  {De~Palo}}, \bibinfo {author} {\bibfnamefont {F.}~\bibnamefont {Rapisarda}},
  \ and\ \bibinfo {author} {\bibfnamefont {G.}~\bibnamefont {Senatore}},\
  }\href {\doibase 10.1103/PhysRevLett.88.206401} {\bibfield  {journal}
  {\bibinfo  {journal} {Phys. Rev. Lett.}\ }\textbf {\bibinfo {volume} {88}},\
  \bibinfo {pages} {206401} (\bibinfo {year} {2002})}\BibitemShut {NoStop}%
\bibitem [{\citenamefont {Maezono}\ \emph {et~al.}(2013)\citenamefont
  {Maezono}, \citenamefont {L\'opez~R\'{\i}os}, \citenamefont {Ogawa},\ and\
  \citenamefont {Needs}}]{PhysRevLett.110.216407}%
  \BibitemOpen
  \bibfield  {author} {\bibinfo {author} {\bibfnamefont {R.}~\bibnamefont
  {Maezono}}, \bibinfo {author} {\bibfnamefont {P.}~\bibnamefont
  {L\'opez~R\'{\i}os}}, \bibinfo {author} {\bibfnamefont {T.}~\bibnamefont
  {Ogawa}}, \ and\ \bibinfo {author} {\bibfnamefont {R.~J.}\ \bibnamefont
  {Needs}},\ }\href {\doibase 10.1103/PhysRevLett.110.216407} {\bibfield
  {journal} {\bibinfo  {journal} {Phys. Rev. Lett.}\ }\textbf {\bibinfo
  {volume} {110}},\ \bibinfo {pages} {216407} (\bibinfo {year}
  {2013})}\BibitemShut {NoStop}%
\bibitem [{\citenamefont {Schindler}\ and\ \citenamefont
  {Zimmermann}(2008)}]{PhysRevB.78.045313}%
  \BibitemOpen
  \bibfield  {author} {\bibinfo {author} {\bibfnamefont {C.}~\bibnamefont
  {Schindler}}\ and\ \bibinfo {author} {\bibfnamefont {R.}~\bibnamefont
  {Zimmermann}},\ }\href {\doibase 10.1103/PhysRevB.78.045313} {\bibfield
  {journal} {\bibinfo  {journal} {Phys. Rev. B}\ }\textbf {\bibinfo {volume}
  {78}},\ \bibinfo {pages} {045313} (\bibinfo {year} {2008})}\BibitemShut
  {NoStop}%
\bibitem [{\citenamefont {Lee}\ \emph {et~al.}(2009)\citenamefont {Lee},
  \citenamefont {Drummond},\ and\ \citenamefont {Needs}}]{PhysRevB.79.125308}%
  \BibitemOpen
  \bibfield  {author} {\bibinfo {author} {\bibfnamefont {R.~M.}\ \bibnamefont
  {Lee}}, \bibinfo {author} {\bibfnamefont {N.~D.}\ \bibnamefont {Drummond}}, \
  and\ \bibinfo {author} {\bibfnamefont {R.~J.}\ \bibnamefont {Needs}},\ }\href
  {\doibase 10.1103/PhysRevB.79.125308} {\bibfield  {journal} {\bibinfo
  {journal} {Phys. Rev. B}\ }\textbf {\bibinfo {volume} {79}},\ \bibinfo
  {pages} {125308} (\bibinfo {year} {2009})}\BibitemShut {NoStop}%
\bibitem [{\citenamefont {Sharma}\ \emph {et~al.}(2018)\citenamefont {Sharma},
  \citenamefont {Saini},\ and\ \citenamefont {Bahuguna}}]{sharma2018phase}%
  \BibitemOpen
  \bibfield  {author} {\bibinfo {author} {\bibfnamefont {R.~O.}\ \bibnamefont
  {Sharma}}, \bibinfo {author} {\bibfnamefont {L.}~\bibnamefont {Saini}}, \
  and\ \bibinfo {author} {\bibfnamefont {B.~P.}\ \bibnamefont {Bahuguna}},\
  }\href@noop {} {\bibfield  {journal} {\bibinfo  {journal} {Journal of
  Physics: Condensed Matter}\ }\textbf {\bibinfo {volume} {30}},\ \bibinfo
  {pages} {185404} (\bibinfo {year} {2018})}\BibitemShut {NoStop}%
\bibitem [{\citenamefont {KuneÅ¡}(2015)}]{0953-8984-27-33-333201}%
  \BibitemOpen
  \bibfield  {author} {\bibinfo {author} {\bibfnamefont {J.}~\bibnamefont
  {KuneÅ¡}},\ }\href {http://stacks.iop.org/0953-8984/27/i=33/a=333201}
  {\bibfield  {journal} {\bibinfo  {journal} {Journal of Physics: Condensed
  Matter}\ }\textbf {\bibinfo {volume} {27}},\ \bibinfo {pages} {333201}
  (\bibinfo {year} {2015})}\BibitemShut {NoStop}%
\bibitem [{\citenamefont {Rademaker}\ \emph
  {et~al.}(2013{\natexlab{a}})\citenamefont {Rademaker}, \citenamefont
  {Johnston}, \citenamefont {Zaanen},\ and\ \citenamefont {van~den
  Brink}}]{PhysRevB.88.235115}%
  \BibitemOpen
  \bibfield  {author} {\bibinfo {author} {\bibfnamefont {L.}~\bibnamefont
  {Rademaker}}, \bibinfo {author} {\bibfnamefont {S.}~\bibnamefont {Johnston}},
  \bibinfo {author} {\bibfnamefont {J.}~\bibnamefont {Zaanen}}, \ and\ \bibinfo
  {author} {\bibfnamefont {J.}~\bibnamefont {van~den Brink}},\ }\href {\doibase
  10.1103/PhysRevB.88.235115} {\bibfield  {journal} {\bibinfo  {journal} {Phys.
  Rev. B}\ }\textbf {\bibinfo {volume} {88}},\ \bibinfo {pages} {235115}
  (\bibinfo {year} {2013}{\natexlab{a}})}\BibitemShut {NoStop}%
\bibitem [{\citenamefont {Rademaker}\ \emph
  {et~al.}(2013{\natexlab{b}})\citenamefont {Rademaker}, \citenamefont {van~den
  Brink}, \citenamefont {Zaanen},\ and\ \citenamefont
  {Hilgenkamp}}]{PhysRevB.88.235127}%
  \BibitemOpen
  \bibfield  {author} {\bibinfo {author} {\bibfnamefont {L.}~\bibnamefont
  {Rademaker}}, \bibinfo {author} {\bibfnamefont {J.}~\bibnamefont {van~den
  Brink}}, \bibinfo {author} {\bibfnamefont {J.}~\bibnamefont {Zaanen}}, \ and\
  \bibinfo {author} {\bibfnamefont {H.}~\bibnamefont {Hilgenkamp}},\ }\href
  {\doibase 10.1103/PhysRevB.88.235127} {\bibfield  {journal} {\bibinfo
  {journal} {Phys. Rev. B}\ }\textbf {\bibinfo {volume} {88}},\ \bibinfo
  {pages} {235127} (\bibinfo {year} {2013}{\natexlab{b}})}\BibitemShut
  {NoStop}%
\bibitem [{\citenamefont {Kaneko}\ \emph {et~al.}(2012)\citenamefont {Kaneko},
  \citenamefont {Seki},\ and\ \citenamefont {Ohta}}]{PhysRevB.85.165135}%
  \BibitemOpen
  \bibfield  {author} {\bibinfo {author} {\bibfnamefont {T.}~\bibnamefont
  {Kaneko}}, \bibinfo {author} {\bibfnamefont {K.}~\bibnamefont {Seki}}, \ and\
  \bibinfo {author} {\bibfnamefont {Y.}~\bibnamefont {Ohta}},\ }\href {\doibase
  10.1103/PhysRevB.85.165135} {\bibfield  {journal} {\bibinfo  {journal} {Phys.
  Rev. B}\ }\textbf {\bibinfo {volume} {85}},\ \bibinfo {pages} {165135}
  (\bibinfo {year} {2012})}\BibitemShut {NoStop}%
\bibitem [{\citenamefont {Kune\ifmmode~\check{s}\else
  \v{s}\fi{}}(2014)}]{PhysRevB.90.235140}%
  \BibitemOpen
  \bibfield  {author} {\bibinfo {author} {\bibfnamefont {J.}~\bibnamefont
  {Kune\ifmmode~\check{s}\else \v{s}\fi{}}},\ }\href {\doibase
  10.1103/PhysRevB.90.235140} {\bibfield  {journal} {\bibinfo  {journal} {Phys.
  Rev. B}\ }\textbf {\bibinfo {volume} {90}},\ \bibinfo {pages} {235140}
  (\bibinfo {year} {2014})}\BibitemShut {NoStop}%
\bibitem [{\citenamefont {Fujiuchi}\ \emph {et~al.}(2018)\citenamefont
  {Fujiuchi}, \citenamefont {Sugimoto},\ and\ \citenamefont
  {Ohta}}]{fujiuchi2018excitonic}%
  \BibitemOpen
  \bibfield  {author} {\bibinfo {author} {\bibfnamefont {R.}~\bibnamefont
  {Fujiuchi}}, \bibinfo {author} {\bibfnamefont {K.}~\bibnamefont {Sugimoto}},
  \ and\ \bibinfo {author} {\bibfnamefont {Y.}~\bibnamefont {Ohta}},\
  }\href@noop {} {\bibfield  {journal} {\bibinfo  {journal} {arXiv preprint
  arXiv:1803.05224}\ } (\bibinfo {year} {2018})}\BibitemShut {NoStop}%
\bibitem [{\citenamefont {Bouadim}\ \emph {et~al.}(2008)\citenamefont
  {Bouadim}, \citenamefont {Batrouni}, \citenamefont {H\'ebert},\ and\
  \citenamefont {Scalettar}}]{PhysRevB.77.144527}%
  \BibitemOpen
  \bibfield  {author} {\bibinfo {author} {\bibfnamefont {K.}~\bibnamefont
  {Bouadim}}, \bibinfo {author} {\bibfnamefont {G.~G.}\ \bibnamefont
  {Batrouni}}, \bibinfo {author} {\bibfnamefont {F.}~\bibnamefont {H\'ebert}},
  \ and\ \bibinfo {author} {\bibfnamefont {R.~T.}\ \bibnamefont {Scalettar}},\
  }\href {\doibase 10.1103/PhysRevB.77.144527} {\bibfield  {journal} {\bibinfo
  {journal} {Phys. Rev. B}\ }\textbf {\bibinfo {volume} {77}},\ \bibinfo
  {pages} {144527} (\bibinfo {year} {2008})}\BibitemShut {NoStop}%
\bibitem [{\citenamefont {Vanhala}\ \emph {et~al.}(2015)\citenamefont
  {Vanhala}, \citenamefont {Baarsma}, \citenamefont {Heikkinen}, \citenamefont
  {Troyer}, \citenamefont {Harju},\ and\ \citenamefont
  {T\"orm\"a}}]{PhysRevB.91.144510}%
  \BibitemOpen
  \bibfield  {author} {\bibinfo {author} {\bibfnamefont {T.~I.}\ \bibnamefont
  {Vanhala}}, \bibinfo {author} {\bibfnamefont {J.~E.}\ \bibnamefont
  {Baarsma}}, \bibinfo {author} {\bibfnamefont {M.~O.~J.}\ \bibnamefont
  {Heikkinen}}, \bibinfo {author} {\bibfnamefont {M.}~\bibnamefont {Troyer}},
  \bibinfo {author} {\bibfnamefont {A.}~\bibnamefont {Harju}}, \ and\ \bibinfo
  {author} {\bibfnamefont {P.}~\bibnamefont {T\"orm\"a}},\ }\href {\doibase
  10.1103/PhysRevB.91.144510} {\bibfield  {journal} {\bibinfo  {journal} {Phys.
  Rev. B}\ }\textbf {\bibinfo {volume} {91}},\ \bibinfo {pages} {144510}
  (\bibinfo {year} {2015})}\BibitemShut {NoStop}%
\bibitem [{\citenamefont {Kaneko}\ \emph {et~al.}(2013)\citenamefont {Kaneko},
  \citenamefont {Ejima}, \citenamefont {Fehske},\ and\ \citenamefont
  {Ohta}}]{PhysRevB.88.035312}%
  \BibitemOpen
  \bibfield  {author} {\bibinfo {author} {\bibfnamefont {T.}~\bibnamefont
  {Kaneko}}, \bibinfo {author} {\bibfnamefont {S.}~\bibnamefont {Ejima}},
  \bibinfo {author} {\bibfnamefont {H.}~\bibnamefont {Fehske}}, \ and\ \bibinfo
  {author} {\bibfnamefont {Y.}~\bibnamefont {Ohta}},\ }\href {\doibase
  10.1103/PhysRevB.88.035312} {\bibfield  {journal} {\bibinfo  {journal} {Phys.
  Rev. B}\ }\textbf {\bibinfo {volume} {88}},\ \bibinfo {pages} {035312}
  (\bibinfo {year} {2013})}\BibitemShut {NoStop}%
\bibitem [{\citenamefont {Zenker}\ \emph {et~al.}(2014)\citenamefont {Zenker},
  \citenamefont {Ihle}, \citenamefont {Bronold},\ and\ \citenamefont
  {Fehske}}]{1742-6596-529-1-012030}%
  \BibitemOpen
  \bibfield  {author} {\bibinfo {author} {\bibfnamefont {B.}~\bibnamefont
  {Zenker}}, \bibinfo {author} {\bibfnamefont {D.}~\bibnamefont {Ihle}},
  \bibinfo {author} {\bibfnamefont {F.~X.}\ \bibnamefont {Bronold}}, \ and\
  \bibinfo {author} {\bibfnamefont {H.}~\bibnamefont {Fehske}},\ }\href
  {http://stacks.iop.org/1742-6596/529/i=1/a=012030} {\bibfield  {journal}
  {\bibinfo  {journal} {Journal of Physics: Conference Series}\ }\textbf
  {\bibinfo {volume} {529}},\ \bibinfo {pages} {012030} (\bibinfo {year}
  {2014})}\BibitemShut {NoStop}%
\bibitem [{\citenamefont {Blankenbecler}\ \emph {et~al.}(1981)\citenamefont
  {Blankenbecler}, \citenamefont {Scalapino},\ and\ \citenamefont
  {Sugar}}]{PhysRevD.24.2278}%
  \BibitemOpen
  \bibfield  {author} {\bibinfo {author} {\bibfnamefont {R.}~\bibnamefont
  {Blankenbecler}}, \bibinfo {author} {\bibfnamefont {D.~J.}\ \bibnamefont
  {Scalapino}}, \ and\ \bibinfo {author} {\bibfnamefont {R.~L.}\ \bibnamefont
  {Sugar}},\ }\href {\doibase 10.1103/PhysRevD.24.2278} {\bibfield  {journal}
  {\bibinfo  {journal} {Phys. Rev. D}\ }\textbf {\bibinfo {volume} {24}},\
  \bibinfo {pages} {2278} (\bibinfo {year} {1981})}\BibitemShut {NoStop}%
\bibitem [{\citenamefont {White}\ \emph {et~al.}(1989)\citenamefont {White},
  \citenamefont {Scalapino}, \citenamefont {Sugar}, \citenamefont {Loh},
  \citenamefont {Gubernatis},\ and\ \citenamefont
  {Scalettar}}]{PhysRevB.40.506}%
  \BibitemOpen
  \bibfield  {author} {\bibinfo {author} {\bibfnamefont {S.~R.}\ \bibnamefont
  {White}}, \bibinfo {author} {\bibfnamefont {D.~J.}\ \bibnamefont
  {Scalapino}}, \bibinfo {author} {\bibfnamefont {R.~L.}\ \bibnamefont
  {Sugar}}, \bibinfo {author} {\bibfnamefont {E.~Y.}\ \bibnamefont {Loh}},
  \bibinfo {author} {\bibfnamefont {J.~E.}\ \bibnamefont {Gubernatis}}, \ and\
  \bibinfo {author} {\bibfnamefont {R.~T.}\ \bibnamefont {Scalettar}},\ }\href
  {\doibase 10.1103/PhysRevB.40.506} {\bibfield  {journal} {\bibinfo  {journal}
  {Phys. Rev. B}\ }\textbf {\bibinfo {volume} {40}},\ \bibinfo {pages} {506}
  (\bibinfo {year} {1989})}\BibitemShut {NoStop}%
\bibitem [{\citenamefont {Santos}(2003)}]{santos2003introduction}%
  \BibitemOpen
  \bibfield  {author} {\bibinfo {author} {\bibfnamefont {R.~R.~d.}\
  \bibnamefont {Santos}},\ }\href@noop {} {\bibfield  {journal} {\bibinfo
  {journal} {Brazilian Journal of Physics}\ }\textbf {\bibinfo {volume} {33}},\
  \bibinfo {pages} {36} (\bibinfo {year} {2003})}\BibitemShut {NoStop}%
\bibitem [{\citenamefont {Koonin}\ \emph {et~al.}(1997)\citenamefont {Koonin},
  \citenamefont {Dean},\ and\ \citenamefont {Langanke}}]{koonin1997shell}%
  \BibitemOpen
  \bibfield  {author} {\bibinfo {author} {\bibfnamefont {S.~E.}\ \bibnamefont
  {Koonin}}, \bibinfo {author} {\bibfnamefont {D.~J.}\ \bibnamefont {Dean}}, \
  and\ \bibinfo {author} {\bibfnamefont {K.}~\bibnamefont {Langanke}},\
  }\href@noop {} {\bibfield  {journal} {\bibinfo  {journal} {Physics reports}\
  }\textbf {\bibinfo {volume} {278}},\ \bibinfo {pages} {1} (\bibinfo {year}
  {1997})}\BibitemShut {NoStop}%
\bibitem [{\citenamefont {Chandrasekharan}\ and\ \citenamefont
  {Wiese}(1999)}]{PhysRevLett.83.3116}%
  \BibitemOpen
  \bibfield  {author} {\bibinfo {author} {\bibfnamefont {S.}~\bibnamefont
  {Chandrasekharan}}\ and\ \bibinfo {author} {\bibfnamefont {U.-J.}\
  \bibnamefont {Wiese}},\ }\href {\doibase 10.1103/PhysRevLett.83.3116}
  {\bibfield  {journal} {\bibinfo  {journal} {Phys. Rev. Lett.}\ }\textbf
  {\bibinfo {volume} {83}},\ \bibinfo {pages} {3116} (\bibinfo {year}
  {1999})}\BibitemShut {NoStop}%
\bibitem [{\citenamefont {Imada}\ and\ \citenamefont
  {Kashima}(2000)}]{imada2000path}%
  \BibitemOpen
  \bibfield  {author} {\bibinfo {author} {\bibfnamefont {M.}~\bibnamefont
  {Imada}}\ and\ \bibinfo {author} {\bibfnamefont {T.}~\bibnamefont
  {Kashima}},\ }\href@noop {} {\bibfield  {journal} {\bibinfo  {journal}
  {Journal of the Physical Society of Japan}\ }\textbf {\bibinfo {volume}
  {69}},\ \bibinfo {pages} {2723} (\bibinfo {year} {2000})}\BibitemShut
  {NoStop}%
\bibitem [{\citenamefont {Wu}\ and\ \citenamefont
  {Zhang}(2005)}]{PhysRevB.71.155115}%
  \BibitemOpen
  \bibfield  {author} {\bibinfo {author} {\bibfnamefont {C.}~\bibnamefont
  {Wu}}\ and\ \bibinfo {author} {\bibfnamefont {S.-C.}\ \bibnamefont {Zhang}},\
  }\href {\doibase 10.1103/PhysRevB.71.155115} {\bibfield  {journal} {\bibinfo
  {journal} {Phys. Rev. B}\ }\textbf {\bibinfo {volume} {71}},\ \bibinfo
  {pages} {155115} (\bibinfo {year} {2005})}\BibitemShut {NoStop}%
\bibitem [{\citenamefont {Berg}\ \emph {et~al.}(2012)\citenamefont {Berg},
  \citenamefont {Metlitski},\ and\ \citenamefont {Sachdev}}]{berg2012sign}%
  \BibitemOpen
  \bibfield  {author} {\bibinfo {author} {\bibfnamefont {E.}~\bibnamefont
  {Berg}}, \bibinfo {author} {\bibfnamefont {M.~A.}\ \bibnamefont {Metlitski}},
  \ and\ \bibinfo {author} {\bibfnamefont {S.}~\bibnamefont {Sachdev}},\
  }\href@noop {} {\bibfield  {journal} {\bibinfo  {journal} {Science}\ }\textbf
  {\bibinfo {volume} {338}},\ \bibinfo {pages} {1606} (\bibinfo {year}
  {2012})}\BibitemShut {NoStop}%
\bibitem [{\citenamefont {Grover}(2013)}]{PhysRevLett.111.130402}%
  \BibitemOpen
  \bibfield  {author} {\bibinfo {author} {\bibfnamefont {T.}~\bibnamefont
  {Grover}},\ }\href {\doibase 10.1103/PhysRevLett.111.130402} {\bibfield
  {journal} {\bibinfo  {journal} {Phys. Rev. Lett.}\ }\textbf {\bibinfo
  {volume} {111}},\ \bibinfo {pages} {130402} (\bibinfo {year}
  {2013})}\BibitemShut {NoStop}%
\bibitem [{\citenamefont {Li}\ \emph {et~al.}(2015)\citenamefont {Li},
  \citenamefont {Jiang},\ and\ \citenamefont {Yao}}]{li2015solving}%
  \BibitemOpen
  \bibfield  {author} {\bibinfo {author} {\bibfnamefont {Z.-X.}\ \bibnamefont
  {Li}}, \bibinfo {author} {\bibfnamefont {Y.-F.}\ \bibnamefont {Jiang}}, \
  and\ \bibinfo {author} {\bibfnamefont {H.}~\bibnamefont {Yao}},\ }\href@noop
  {} {\bibfield  {journal} {\bibinfo  {journal} {Physical Review B}\ }\textbf
  {\bibinfo {volume} {91}},\ \bibinfo {pages} {241117(R)} (\bibinfo {year}
  {2015})}\BibitemShut {NoStop}%
\bibitem [{\citenamefont {Wei}\ \emph {et~al.}(2016)\citenamefont {Wei},
  \citenamefont {Wu}, \citenamefont {Li}, \citenamefont {Zhang},\ and\
  \citenamefont {Xiang}}]{PhysRevLett.116.250601}%
  \BibitemOpen
  \bibfield  {author} {\bibinfo {author} {\bibfnamefont {Z.~C.}\ \bibnamefont
  {Wei}}, \bibinfo {author} {\bibfnamefont {C.}~\bibnamefont {Wu}}, \bibinfo
  {author} {\bibfnamefont {Y.}~\bibnamefont {Li}}, \bibinfo {author}
  {\bibfnamefont {S.}~\bibnamefont {Zhang}}, \ and\ \bibinfo {author}
  {\bibfnamefont {T.}~\bibnamefont {Xiang}},\ }\href {\doibase
  10.1103/PhysRevLett.116.250601} {\bibfield  {journal} {\bibinfo  {journal}
  {Phys. Rev. Lett.}\ }\textbf {\bibinfo {volume} {116}},\ \bibinfo {pages}
  {250601} (\bibinfo {year} {2016})}\BibitemShut {NoStop}%
\bibitem [{\citenamefont {Azaria}\ \emph {et~al.}(2000)\citenamefont {Azaria},
  \citenamefont {Boulat},\ and\ \citenamefont
  {Lecheminant}}]{PhysRevB.61.12112}%
  \BibitemOpen
  \bibfield  {author} {\bibinfo {author} {\bibfnamefont {P.}~\bibnamefont
  {Azaria}}, \bibinfo {author} {\bibfnamefont {E.}~\bibnamefont {Boulat}}, \
  and\ \bibinfo {author} {\bibfnamefont {P.}~\bibnamefont {Lecheminant}},\
  }\href {\doibase 10.1103/PhysRevB.61.12112} {\bibfield  {journal} {\bibinfo
  {journal} {Phys. Rev. B}\ }\textbf {\bibinfo {volume} {61}},\ \bibinfo
  {pages} {12112} (\bibinfo {year} {2000})}\BibitemShut {NoStop}%
\bibitem [{\citenamefont {Assaraf}\ \emph {et~al.}(2004)\citenamefont
  {Assaraf}, \citenamefont {Azaria}, \citenamefont {Boulat}, \citenamefont
  {Caffarel},\ and\ \citenamefont {Lecheminant}}]{PhysRevLett.93.016407}%
  \BibitemOpen
  \bibfield  {author} {\bibinfo {author} {\bibfnamefont {R.}~\bibnamefont
  {Assaraf}}, \bibinfo {author} {\bibfnamefont {P.}~\bibnamefont {Azaria}},
  \bibinfo {author} {\bibfnamefont {E.}~\bibnamefont {Boulat}}, \bibinfo
  {author} {\bibfnamefont {M.}~\bibnamefont {Caffarel}}, \ and\ \bibinfo
  {author} {\bibfnamefont {P.}~\bibnamefont {Lecheminant}},\ }\href {\doibase
  10.1103/PhysRevLett.93.016407} {\bibfield  {journal} {\bibinfo  {journal}
  {Phys. Rev. Lett.}\ }\textbf {\bibinfo {volume} {93}},\ \bibinfo {pages}
  {016407} (\bibinfo {year} {2004})}\BibitemShut {NoStop}%
\bibitem [{\citenamefont {Mermin}\ and\ \citenamefont
  {Wagner}(1966)}]{PhysRevLett.17.1133}%
  \BibitemOpen
  \bibfield  {author} {\bibinfo {author} {\bibfnamefont {N.~D.}\ \bibnamefont
  {Mermin}}\ and\ \bibinfo {author} {\bibfnamefont {H.}~\bibnamefont
  {Wagner}},\ }\href {\doibase 10.1103/PhysRevLett.17.1133} {\bibfield
  {journal} {\bibinfo  {journal} {Phys. Rev. Lett.}\ }\textbf {\bibinfo
  {volume} {17}},\ \bibinfo {pages} {1133} (\bibinfo {year}
  {1966})}\BibitemShut {NoStop}%
\bibitem [{\citenamefont {Berezinskii}(1972)}]{berezinskii1972overview}%
  \BibitemOpen
  \bibfield  {author} {\bibinfo {author} {\bibfnamefont {V.}~\bibnamefont
  {Berezinskii}},\ }\href@noop {} {\bibfield  {journal} {\bibinfo  {journal}
  {Sov. Phys. JETP}\ }\textbf {\bibinfo {volume} {34}},\ \bibinfo {pages} {610}
  (\bibinfo {year} {1972})}\BibitemShut {NoStop}%
\bibitem [{\citenamefont {Kosterlitz}\ and\ \citenamefont
  {Thouless}(1973)}]{Kosterlitz_1973}%
  \BibitemOpen
  \bibfield  {author} {\bibinfo {author} {\bibfnamefont {J.~M.}\ \bibnamefont
  {Kosterlitz}}\ and\ \bibinfo {author} {\bibfnamefont {D.~J.}\ \bibnamefont
  {Thouless}},\ }\href {\doibase 10.1088/0022-3719/6/7/010} {\bibfield
  {journal} {\bibinfo  {journal} {Journal of Physics C: Solid State Physics}\
  }\textbf {\bibinfo {volume} {6}},\ \bibinfo {pages} {1181} (\bibinfo {year}
  {1973})}\BibitemShut {NoStop}%
\bibitem [{\citenamefont {Kosterlitz}(1974)}]{kosterlitz1974critical}%
  \BibitemOpen
  \bibfield  {author} {\bibinfo {author} {\bibfnamefont {J.}~\bibnamefont
  {Kosterlitz}},\ }\href@noop {} {\bibfield  {journal} {\bibinfo  {journal}
  {Journal of Physics C: Solid State Physics}\ }\textbf {\bibinfo {volume}
  {7}},\ \bibinfo {pages} {1046} (\bibinfo {year} {1974})}\BibitemShut
  {NoStop}%
\bibitem [{\citenamefont {Scalapino}\ \emph {et~al.}(1993)\citenamefont
  {Scalapino}, \citenamefont {White},\ and\ \citenamefont
  {Zhang}}]{scalapino1993insulator}%
  \BibitemOpen
  \bibfield  {author} {\bibinfo {author} {\bibfnamefont {D.~J.}\ \bibnamefont
  {Scalapino}}, \bibinfo {author} {\bibfnamefont {S.~R.}\ \bibnamefont
  {White}}, \ and\ \bibinfo {author} {\bibfnamefont {S.}~\bibnamefont
  {Zhang}},\ }\href@noop {} {\bibfield  {journal} {\bibinfo  {journal}
  {Physical Review B}\ }\textbf {\bibinfo {volume} {47}},\ \bibinfo {pages}
  {7995} (\bibinfo {year} {1993})}\BibitemShut {NoStop}%
\bibitem [{\citenamefont {Xu}\ \emph {et~al.}(2017)\citenamefont {Xu},
  \citenamefont {Sun}, \citenamefont {Schattner}, \citenamefont {Berg},\ and\
  \citenamefont {Meng}}]{xu2017non}%
  \BibitemOpen
  \bibfield  {author} {\bibinfo {author} {\bibfnamefont {X.~Y.}\ \bibnamefont
  {Xu}}, \bibinfo {author} {\bibfnamefont {K.}~\bibnamefont {Sun}}, \bibinfo
  {author} {\bibfnamefont {Y.}~\bibnamefont {Schattner}}, \bibinfo {author}
  {\bibfnamefont {E.}~\bibnamefont {Berg}}, \ and\ \bibinfo {author}
  {\bibfnamefont {Z.~Y.}\ \bibnamefont {Meng}},\ }\href@noop {} {\bibfield
  {journal} {\bibinfo  {journal} {Physical Review X}\ }\textbf {\bibinfo
  {volume} {7}},\ \bibinfo {pages} {031058} (\bibinfo {year}
  {2017})}\BibitemShut {NoStop}%
\bibitem [{\citenamefont {Paiva}\ \emph {et~al.}(2004)\citenamefont {Paiva},
  \citenamefont {dos Santos}, \citenamefont {Scalettar},\ and\ \citenamefont
  {Denteneer}}]{PhysRevB.69.184501}%
  \BibitemOpen
  \bibfield  {author} {\bibinfo {author} {\bibfnamefont {T.}~\bibnamefont
  {Paiva}}, \bibinfo {author} {\bibfnamefont {R.~R.}\ \bibnamefont {dos
  Santos}}, \bibinfo {author} {\bibfnamefont {R.~T.}\ \bibnamefont
  {Scalettar}}, \ and\ \bibinfo {author} {\bibfnamefont {P.~J.~H.}\
  \bibnamefont {Denteneer}},\ }\href {\doibase 10.1103/PhysRevB.69.184501}
  {\bibfield  {journal} {\bibinfo  {journal} {Phys. Rev. B}\ }\textbf {\bibinfo
  {volume} {69}},\ \bibinfo {pages} {184501} (\bibinfo {year}
  {2004})}\BibitemShut {NoStop}%
\bibitem [{\citenamefont {Nelson}\ and\ \citenamefont
  {Kosterlitz}(1977)}]{PhysRevLett.39.1201}%
  \BibitemOpen
  \bibfield  {author} {\bibinfo {author} {\bibfnamefont {D.~R.}\ \bibnamefont
  {Nelson}}\ and\ \bibinfo {author} {\bibfnamefont {J.~M.}\ \bibnamefont
  {Kosterlitz}},\ }\href {\doibase 10.1103/PhysRevLett.39.1201} {\bibfield
  {journal} {\bibinfo  {journal} {Phys. Rev. Lett.}\ }\textbf {\bibinfo
  {volume} {39}},\ \bibinfo {pages} {1201} (\bibinfo {year}
  {1977})}\BibitemShut {NoStop}%
\bibitem [{\citenamefont {Berche}\ \emph {et~al.}(2002)\citenamefont {Berche},
  \citenamefont {Sanchez} \emph {et~al.}}]{berche2002correlations}%
  \BibitemOpen
  \bibfield  {author} {\bibinfo {author} {\bibfnamefont {B.}~\bibnamefont
  {Berche}}, \bibinfo {author} {\bibfnamefont {A.~F.}\ \bibnamefont {Sanchez}},
   \emph {et~al.},\ }\href@noop {} {\bibfield  {journal} {\bibinfo  {journal}
  {EPL (Europhysics Letters)}\ }\textbf {\bibinfo {volume} {60}},\ \bibinfo
  {pages} {539} (\bibinfo {year} {2002})}\BibitemShut {NoStop}%
\bibitem [{\citenamefont {Moreo}\ and\ \citenamefont
  {Scalapino}(1991)}]{PhysRevLett.66.946}%
  \BibitemOpen
  \bibfield  {author} {\bibinfo {author} {\bibfnamefont {A.}~\bibnamefont
  {Moreo}}\ and\ \bibinfo {author} {\bibfnamefont {D.~J.}\ \bibnamefont
  {Scalapino}},\ }\href {\doibase 10.1103/PhysRevLett.66.946} {\bibfield
  {journal} {\bibinfo  {journal} {Phys. Rev. Lett.}\ }\textbf {\bibinfo
  {volume} {66}},\ \bibinfo {pages} {946} (\bibinfo {year} {1991})}\BibitemShut
  {NoStop}%
\bibitem [{\citenamefont {Kohno}\ and\ \citenamefont
  {Takahashi}(1997)}]{PhysRevB.56.3212}%
  \BibitemOpen
  \bibfield  {author} {\bibinfo {author} {\bibfnamefont {M.}~\bibnamefont
  {Kohno}}\ and\ \bibinfo {author} {\bibfnamefont {M.}~\bibnamefont
  {Takahashi}},\ }\href {\doibase 10.1103/PhysRevB.56.3212} {\bibfield
  {journal} {\bibinfo  {journal} {Phys. Rev. B}\ }\textbf {\bibinfo {volume}
  {56}},\ \bibinfo {pages} {3212} (\bibinfo {year} {1997})}\BibitemShut
  {NoStop}%
\bibitem [{\citenamefont {Batrouni}\ and\ \citenamefont
  {Scalettar}(2000)}]{PhysRevLett.84.1599}%
  \BibitemOpen
  \bibfield  {author} {\bibinfo {author} {\bibfnamefont {G.~G.}\ \bibnamefont
  {Batrouni}}\ and\ \bibinfo {author} {\bibfnamefont {R.~T.}\ \bibnamefont
  {Scalettar}},\ }\href {\doibase 10.1103/PhysRevLett.84.1599} {\bibfield
  {journal} {\bibinfo  {journal} {Phys. Rev. Lett.}\ }\textbf {\bibinfo
  {volume} {84}},\ \bibinfo {pages} {1599} (\bibinfo {year}
  {2000})}\BibitemShut {NoStop}%
\bibitem [{\citenamefont {Schmid}\ \emph {et~al.}(2002)\citenamefont {Schmid},
  \citenamefont {Todo}, \citenamefont {Troyer},\ and\ \citenamefont
  {Dorneich}}]{PhysRevLett.88.167208}%
  \BibitemOpen
  \bibfield  {author} {\bibinfo {author} {\bibfnamefont {G.}~\bibnamefont
  {Schmid}}, \bibinfo {author} {\bibfnamefont {S.}~\bibnamefont {Todo}},
  \bibinfo {author} {\bibfnamefont {M.}~\bibnamefont {Troyer}}, \ and\ \bibinfo
  {author} {\bibfnamefont {A.}~\bibnamefont {Dorneich}},\ }\href {\doibase
  10.1103/PhysRevLett.88.167208} {\bibfield  {journal} {\bibinfo  {journal}
  {Phys. Rev. Lett.}\ }\textbf {\bibinfo {volume} {88}},\ \bibinfo {pages}
  {167208} (\bibinfo {year} {2002})}\BibitemShut {NoStop}%
\bibitem [{\citenamefont {Yang}\ and\ \citenamefont
  {Feiguin}(2018)}]{PhysRevB.98.035128}%
  \BibitemOpen
  \bibfield  {author} {\bibinfo {author} {\bibfnamefont {C.}~\bibnamefont
  {Yang}}\ and\ \bibinfo {author} {\bibfnamefont {A.~E.}\ \bibnamefont
  {Feiguin}},\ }\href {\doibase 10.1103/PhysRevB.98.035128} {\bibfield
  {journal} {\bibinfo  {journal} {Phys. Rev. B}\ }\textbf {\bibinfo {volume}
  {98}},\ \bibinfo {pages} {035128} (\bibinfo {year} {2018})}\BibitemShut
  {NoStop}%
\bibitem [{\citenamefont {Jiang}\ \emph {et~al.}(2017)\citenamefont {Jiang},
  \citenamefont {Li}, \citenamefont {Kivelson},\ and\ \citenamefont
  {Yao}}]{PhysRevB.95.241103}%
  \BibitemOpen
  \bibfield  {author} {\bibinfo {author} {\bibfnamefont {Y.-F.}\ \bibnamefont
  {Jiang}}, \bibinfo {author} {\bibfnamefont {Z.-X.}\ \bibnamefont {Li}},
  \bibinfo {author} {\bibfnamefont {S.~A.}\ \bibnamefont {Kivelson}}, \ and\
  \bibinfo {author} {\bibfnamefont {H.}~\bibnamefont {Yao}},\ }\href {\doibase
  10.1103/PhysRevB.95.241103} {\bibfield  {journal} {\bibinfo  {journal} {Phys.
  Rev. B}\ }\textbf {\bibinfo {volume} {95}},\ \bibinfo {pages} {241103(R)}
  (\bibinfo {year} {2017})}\BibitemShut {NoStop}%
\end{thebibliography}%

\end{document}